# Continuous-feed nanocasting process for the synthesis of bismuth nanowire composites

K. Vandaele[1], J. P. Heremans[2,3,4], I. Van Driessche[1], P. Van Der Voort[1], and K. De Buysser[1]

**We present a novel, continuous-feed nanocasting procedure for the synthesis of bismuth nanowire structures embedded in the pores of a mesoporous silica template. The immobilization of a bismuth salt inside the silica template from a diluted metal salt solution yields a sufficiently high loading to obtain electrically conducting bulk nanowire composite samples after reduction and sintering the nanocomposite powders. Electrical resistivity measurements of sintered bismuth nanowires embedded in the silica template reveal size-quantization effects.**

Thermoelectrics could play an important role in waste heat recovery and solid-state cooling as they possess the ability to convert heat directly to electricity and vice versa.[1,2] The advantages of thermoelectrics include quiet operation, no mechanical moving parts, and a long lifetime. However, the limiting factor preventing the large-scale production of thermoelectrics for both power production and solid-state cooling is their low efficiency. A material's efficiency is described by the figure of merit, denoted $zT$, and is defined as: $zT = S^2T/\rho\kappa$, with $S$ the Seebeck coefficient, $\rho$ the electrical resistivity, and $\kappa$ the thermal conductivity.

Pioneering work by Hicks and Dresselhaus[3] theoretically predicted that nanostructured thermoelectrics, could have an enhanced $zT$ value compared to their bulk counterparts[3,4]. Such increase of $zT$ was predicted in cylindrical Bi nanowires by Lin et al.[5] The synthesis of Bi nanowires by pressure injection[6] of a melt in porous, anodized alumina was reported Zhang et al., while Heremans et al. reported the synthesis of Bi nanowires by the impregnation of Bi vapour in an alumina template[7]. Attempts at making Bi nanowire thermoelectrics[7] have resulted in large thermopowers, but comparatively small $zT$s. One issue has been that no $Bi_{1-x}Sb_x$ nanowires could be prepared by conventional methods due the large difference in melting point and vapour pressure between Bi and Sb, whereas theory[8] predicted superior performance in these alloy wires. In addition, the high thermal conductivity of an alumina matrix causes parasitic thermal losses, deteriorating the $zT$.

A suitable chemical route to synthesize non-siliceous nanostructured and mesoporous materials is by means of a hard template replication method, known as nanocasting.[9,10] Typically, the pores of mesoporous silica are used as a nano-mold or template. Since mesoporous silica materials possess pores with dimensions ranging from approximately 2 to 50 nm, they are perfect as a template material to synthesize Bi nanowire composites with diameters well below the threshold of 50 nm to observe a semimetal-to-semiconductor transition.[5]

A crucial step in the nanocasting replication process is the impregnation of the precursor material inside the template´s pores and the conversion of that precursor to the target material. Typically, a metal salt dissolved in ethanol or water is loaded inside the mesoporous template through conventional solvent

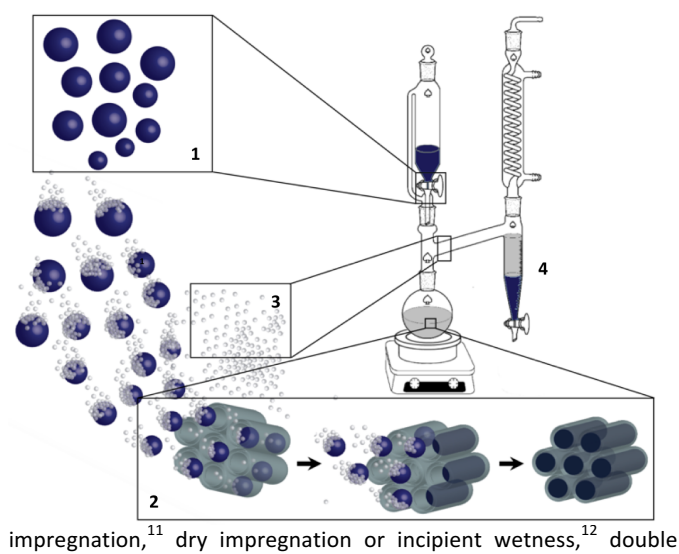

impregnation,[11] dry impregnation or incipient wetness,[12] double

Figure 1. Schematic of the continuous-feed impregnation procedure in refluxing n-octane (bp. 125°C). (1) The aqueous precursor solution is added to the silica/n-octane dispersion using an addition funnel or syringe pump. (2) The polar precursor solution and n-octane from a two-phase system. Meanwhile, the precursor infiltrates the silica template through capillary impregnation. (3) As the impregnation is performed in refluxing n-octane, the water boils off and the diluted precursor solution inside the pores becomes more concentrated until all water is removed and only the precursor salt remains. Partial decomposition of the precursor salt may occur. (4) The aqueous solution is collected by means of a Dean-Stark separator.

[1] Department of Inorganic and Physical Chemistry, Ghent University, Krijgslaan 281 - S3, 9000 Gent, Belgium.
[2] Department of Mechanical and Aerospace Engineering, Columbus, Ohio, 43210, USA
[3] Department of Physics, The Ohio State University, Columbus, Ohio, 43210, USA
[4] Department of Materials Science and Engineering, The Ohio State University, Columbus, Ohio, 43210, USA

solvent,[13,14] or evaporation-induced impregnation method, and subsequently subjected to a thermal treatment.[9,10,15,16,17] The replicated mesoporous material is obtained after template removal through chemical etching with NaOH or HF. However, problems associated with the current impregnation methods of mesoporous templates include the low filling degree, inhomogeneous filling of the pores, or the deposition of precursor material on the external surface of the porous template. In many cases, multiple loading steps are required to obtain an acceptable degree of loading.[11,18,19] Ordered, mesoporous silica templates were used by Xu et al. for the synthesis of Bi nanowires by the decomposition of triphenylbismuth in supercritical toluene.[20] However, only small quantities of material were obtained.

In the present work, a scalable, continuous-feed impregnation process based on the two-solvent impregnation method[21] was used. This process was developed to fill the pores of a mesoporous silica template with bismuth metal salt by impregnating the template with a diluted acidic precursor solution. A low-temperature reduction treatment (220 to 230 °C) allowed the conversion of the impregnated bismuth salt to metallic bismuth nanowires confined in the template's pores. Powder processing techniques allowed the synthesis of bulk-sized samples composed of Bi nanowires, referred to as bulk nanowire composites. Electrical resistivity measurements were performed to identify size quantization effects in the bulk nanowire composite.

The two-solvent impregnation method[21] (Figure ESI 3) consists of the addition of precursor solution in an amount equal to the pore volume of the template in a dispersion of the template in hexane. In the process described here, the precursor solution is added continuously to a dispersion of the mesoporous silica template in refluxing toluene or n-octane. By means of a Dean-Stark separator, the aqueous phase is eliminated from the system, leading to the immobilization of the metal precursor salt inside the template's pores. All chemicals are listed in the electronic supplementary information (ESI section 1).

In a typical synthesis, 1 g SBA-15 or KIT-6 mesoporous silica (ESI section 3) with a pore volume of 1 $cm^3$/g was dispersed in 100 mL n-octane in a 250 mL fluoropolymer (PFA) round bottom flask (ESI section 4). The amount of metal salt required to fill the template's pores was derived based on the template's pore volume and the density of the precursor salt. For the impregnation of $BiCl_3$, the precursor solution was prepared by dissolving 3.5 g $Bi_2O_3$ in 12 mL 36 w% HCl and 24 mL $H_2O$. The precursor solution was added to the dispersion of silica in n-octane at a rate of 4 mL/h using a syringe pump, while the non-polar solvent was refluxing (Figure ESI 4). All studied samples are listed in Table ESI 1. The impregnated silica powder was collected through filtration and subsequently reduced in a sealable flow furnace at 230 °C for 12 h in a $N_2$ - 5% $H_2$ gas flow mixed with hydrazine vapour. The hydrazine vapour was carried through the furnace by bubbling the $N_2$ - 5% $H_2$ gas through a hydrazine monohydrate solution (Figure ESI 7). The Bi – $SiO_2$ nanowire composite powder was removed from the flow furnace inside an argon glovebox, and approximately 2 g sample was transferred to a 10 mm Ø graphite die, which was sealed with rubber glue to avoid oxidation when transferred to the spark plasma sintering (SPS) device. The nanowire composite powder was sintered at 230 °C for 20 min in a vacuum under a uniaxial pressure of 50 MPa. The sample preparations for four-probe electrical transport measurements and mounting inside the cryostat were performed in an argon glovebox. Typical sample dimensions for electrical transport measurements were 1.5 x 1 x 7 mm. For $N_2$ sorption analysis, the silica template of the reduced Bi – $SiO_2$ powder was dissolved in a 1 mol/L NaOH solution for 3 h. The mesoporous Bi powder was isolated by centrifugation and washed four times with water.

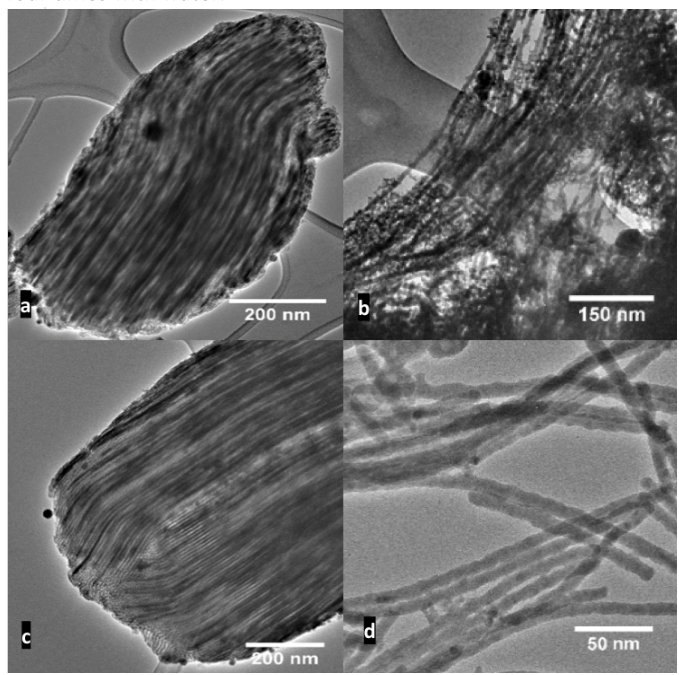

Figure 2. (a) TEM image of SBA-15 filled with 12 v% Bi, prepared by the impregnation of $Bi(NO_3)_3 \cdot 5H_2O$ dissolved in 5 mol/L $HNO_3$, with toluene as non-polar solvent and reduced at 220 °C in a hydrazine loaded Ar-5% $H_2$ gas flow for 12 h. (b) Bi nanowires after chemical etching of the silica template of the sample in 1 mol/L NaOH solution for 3 h. (c) TEM image of SBA-15 filled with 30 v% Bi, prepared by the impregnation of $BiCl_3$ dissolved in a 4 mol/L HCl – 80 v% MeOH solution, with n-octane as non-polar solvent and reduced at 220 °C in a hydrazine loaded Ar-5% $H_2$ gas flow for 12 h. (d) Bi nanowires after chemical etching of sample c. Note that Bi nanowires oxidize spontaneous in air, therefore the analysed samples referred to as Bi are oxidized.

The X-ray powder diffraction (XRPD) patterns in Figure ESI 10 show that the bismuth precursor salt reduced to bismuth at 230 °C for 12 h in the presence of hydrazine vapour. When measured under argon atmosphere, no secondary phases were observed. However, upon removing the tape, which sealed the sample from air, the XRPD pattern indicates the presence of $Bi_2O_{2.5}$, which formed within seconds. KIT-6 mesoporous silica, with a 3D interconnected mesopore system, was used as template for the preparation of mesoporous bismuth powder for $N_2$ physisorption measurements and for electric transport measurements. Mesoporous bismuth, was obtained after chemical etching of the KIT-6 silica template. A BET surface area of 20 $m^2$/g and 13 $m^2$/g was obtained when $Bi(NO_3)_3$ and $BiCl_3$, respectively, were used as precursor salts (Figure ESI 11). The pore volumes were 0.14 and 0.06 $cm^3$/g, respectively.

SBA-15 mesoporous silica, which possesses hexagonally packed, cylindrical mesopores, was used as template for the TEM analysis shown in Figure 2. The TEM images depict oxidized bismuth nanowires (due to sample preparation for TEM) with a diameter between 5 and 10 nm confined in the template's pores (a and c) and the replicated nanowires after etching the template (b and d). The appearance of the nanowires depends on the precursor salt used. Namely, for $Bi(NO_3)_3$, rough nanowires were obtained, whereas smooth wires were obtained for $BiCl_3$. The latter also forms a melt at approximately 228 °C according to DTA shown in Figure ESI 6. The theoretical volume fractions of Bi confined in the pores are 12 v% and 30 v% for $Bi(NO_3)_3$ and $BiCl_3$, respectively.

Figure 3 depicts the carrier concentration of a sintered Bi – SiO$_2$ (KIT-6) nanowire composite sample as function of *1/2kT*. By fitting the carrier concentration for an intrinsic semiconductor through the data points, a band gap of 45 meV was calculated. Figure 4 compares the electrical resistivity of bulk Bi with that of Bi nanowire composites normalized by their value at 300 K. The data shows the occurrence of size-quantization effects in the nanowire composite samples, since the electrical resistivity behaviour follows a semiconducting behaviour instead of a metallic one, as in bulk Bi.

The loading of the pores changes significantly depending on the precursor salt used, given the difference in density and molecular weight of Bi(NO$_3$)$_3$ and BiCl$_3$. If we consider complete filling of the pores with the precursor salts, 12 v% and 30 v%, Bi can be deposited in the pores after reduction of Bi(NO$_3$)$_3$ and BiCl$_3$, respectively. How the loading could be further enhanced is shown in Table ESI 2. KIT-6, a mesoporous silica template with an interconnected pore system, was used for the synthesis of mesoporous Bi. No high-BET surface area was obtained for either sample, which primarily is due to the high density of Bi (9.78 g/cm$^3$) and its low melting point (271.5 °C). Bismuth also can leach out of a template's pores during reduction of the bismuth salt. This was mostly observed for BiCl$_3$, as it also forms a melt at approximately 228 °C (Figure ESI 6). We believe that the smooth nanowires in Figure 2 were obtained specifically because of the formation that melt. The use of BiCl$_3$ as precursor was preferred since it can lead to a higher loading of Bi inside the template. This is particularly important for the synthesis of nanocomposites, as a sufficiently high loading of Bi in the template is required to form an electrical percolation path when the Bi – SiO$_2$ nanowire composite powder were sintered. For the same reason, the use of a silica template with 3D interconnected pore system (KIT-6) is preferred over a template with linear pores (SBA-15). Electric transport measurements were performed on a bulk nanowire composite sample using KIT-6 with a pore diameter of approximately 7 nm as silica template. Based on the amount of BiCl$_3$ precursor salt impregnated in the template, the estimated composition of the composite sample was 33 v% silica, 29 v% Bi nanowires and 38 v% voids.

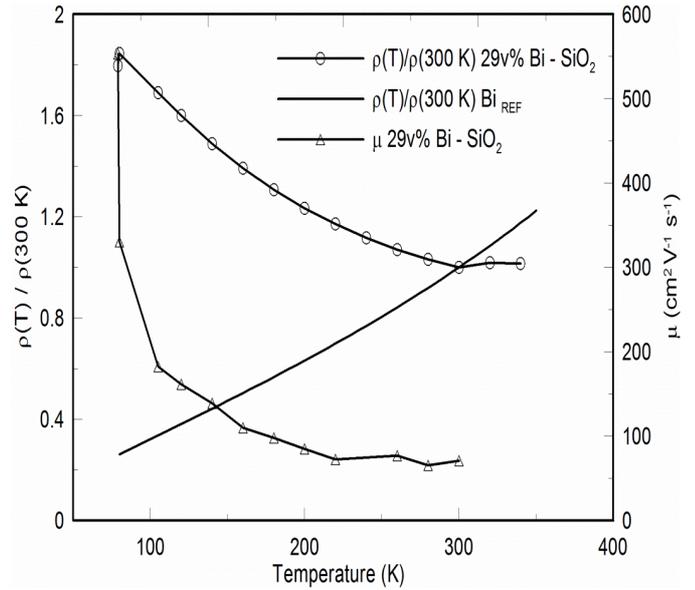

Figure 4. Electrical resistivity of 29 v% Bi – SiO$_2$ nanowire composite sample and bulk bismuth normalized by their value at 300 K. Where the normalized resistivity of bulk Bi drops, and thus exhibits a metallic character when cooled down to 77 K, the relative resistivity of 29 v% Bi – SiO$_2$ increases, providing evidence for size-quantization effects in sintered bismuth nanowire composites. The carrier mobility is depicted on the right axis. Note that all sample preparations and measurements were performed under inert atmosphere.

Evidence of size quantization in the bismuth nanowire composites was obtained from measurements of the Hall effect (Figure 3) and resistivity (Figure 4) of the sintered samples. For the nanocomposites, the Hall resistivity is corrected for the volume fractions of Bi by using the effective medium theory[22] for the Hall effect; for approx. 30% Bi filling by volume, that correction is a factor of 2.2, does not affect the T-dependence nor the band gap. The inverse of the low-field (< 1 Tesla) Hall slope $R_H$ was interpreted to give the charge-carrier concentration via $n = (e\ R_H)^{-1}$ with $e$ the electron charge. $R_H$ was positive, so the dominant carriers are holes. Figure 3 shows the temperature dependence of $n$ measured on the sample (29 v% Bi – SiO$_2$), in an Arrhenius plot. A very clean, activated behavior is observed, following a law $n(T) = n_0\ e^{-E_g/2k_BT}$ valid for intrinsic semiconductors, where $E_g$ is the band gap of the semiconductor. A mixture of very narrow nanowires with bulk-like Bi (which displays a metal-like conductivity) will not generate such a simple temperature-dependence. Furthermore, the charge-carrier concentration measured on the nanowire sample is an order of magnitude smaller than that (also shown in Figure 3) of bulk Bi.[23] Finally, the value of $E_g$ = 45 meV obtained from the Arrhenius plot in Figure 3 corresponds very well to the theoretical value of the band gap opened by size-quantization effects in Bi nanowires of about 20 nm diameter, providing evidence of size quantization effects in the nanowire composites.

At first sight, the 20 nm diameter measured by electrical measurements appears to contradict the observation that the template's pores were 7 nm, which should correspond to a bandgap exceeding 100 meV. However, it has been observed experimentally[7] that conduction in nanowires of diameter smaller than 9 nm becomes localized. In contrast, no localization is observed in the sample measured here, as is illustrated by the resistivity data in Figure 4. These data show a temperature dependence consistent with the decrease in carrier concentration, very different from bulk Bi (also shown) and conducive to mobilities (shown using the right-hand ordinate axes) of the order of 500 to 50 cm$^2$/V sec. Because there is always a statistical distribution of nanowire diameter, these observations lead to the conclusion that the conduction is dominated by nanowires with diameters near 20 nm with conduction in all the narrower wires simply localized. Indeed, there is a lower limit to the range of validity of the theoretical calculations by Lin et al.[5] set by the condition for Anerson localization[24]: the product $k_F·l > 1$, where $k_F$ is the Fermi wavevector and $l$ is the electronic mean-free path, here limited by the wire diameter. For bulk Bi, $k_F$ is given in Ref. 25 to be very anisotropic with the smallest value 1 x 10$^8$ m$^{-1}$, meaning that nanowires with diameters below 10 nm should not conduct. Because of quantization, the carrier concentration in the present sample is smaller still, leading to a smaller $k_F$; therefore, the minimum wire diameter must be larger, e.g. on the order of the observed 20 nm.

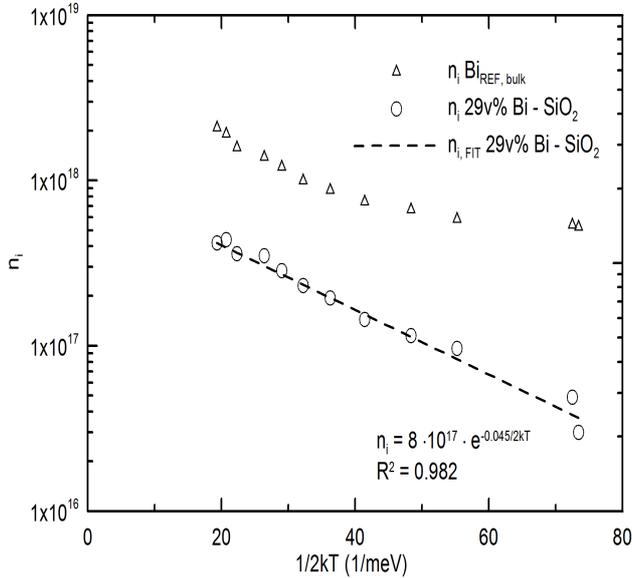

Figure 3. Carrier concentration of bulk bismuth and 29 v% Bi – SiO$_2$ (KIT-6) nanowire composite sample as function of 1/2kT. The lower carrier concentration of the nanocomposite sample compared to bulk Bi can be attributed to size-quantization effects in the nanowires. The carrier concentration for the intrinsic semiconductor was derived by $n_i = n_0\ e^{-E_g/2kT}$. The band gap obtained by the fit through the data points yields: $E_g$ = 45 meV, which corresponds to Bi nanowires with diameter 20 nm according to Lin et al.[5] Note that all sample preparations and measurements were performed under inert atmosphere.

In conclusion, a facile nanocasting technique was developed for the synthesis of bismuth nanowire composites. By means of the continuous-feed nanocasting process and a low-temperature reduction treatment of the impregnated bismuth salt, we were able to synthesize bismuth nanowires embedded in the pores of a mesoporous silica powder. The nanocasting method enabled a sufficiently high loading of bismuth inside the pores of a 3D interconnected silica template to form electrically conducting nanowire composite samples. This provides the opportunity to fabricate a bulk material composed of nanowires with potentially enhanced thermoelectric properties. Evidence of size-quantization effects was obtained from electrical resistivity and carrier concentration data, which show the clear signature of semiconducting band conduction with an energy gap of 45 meV and no signs of localization. The results suggest several directions for further research, e.g. aliovalent doping and alloying Bi with Sb, as well as measurements of the thermoelectric properties of the material, in order to enhance the material's $zT$.

## Acknowledgements

The authors would like to acknowledge funding of this research by the Flemish Government (Innovation Science and Technology project) and the Flanders Research Foundation (FWO). KV would like to thank Bin He and Prof. J.P. Heremans for many fruitful discussions, and in particular, thanks to Katrien Haustraete and Katrien De Keukeleere for all TEM measurements. KV also thanks María Dolores González Gómez for her support throughout the research and for making the abstract figure.

## Notes and references

**Electronic Supplementary Information**

Continuous-feed nanocasting process for the synthesis of bismuth nanowire composites

*1    Materials*

Pluronics P123 ($EO_{20}PO_{70}EO_{20}$, Mw = 5800), 98% reagent grade tetraethylorthosilicate (TEOS), 65w% nitric acid, 36w% hydrochloric acid, 98.5 % xylenes and 99% n-butanol were purchased from Sigma Aldrich. 99.999% bismuth(III) oxide and 98% hydrazine monohydrate were supplied by Alfa Aesar. 99% toluene and 98% methanol were supplied by Fiers, while 97% n-octane was purchased from TCI Europe N.V. All reagents were used as received.

*2    Characterization*

X-Ray powder Diffraction (XRD) patterns were recorded on a Thermo Scientific ARL X'TRA X-Ray Diffractometer with the Bragg–Brentano theta-2 theta configuration and using Cu Ka radiation. Topas Academic V4.1 software was used for Rietveld refinement[1]. Nitrogen sorption experiments were performed at 77 K with a Micromeritics TriStar 3000 device. Samples were vacuum dried at 120 °C for 12 h prior to analysis. The surface area was calculated using the BET method while the pore-size distribution was determined by analysis of the adsorption branch of the isotherms using the BJH method. Thermogravimetric analysis (TGA) was performed on a NETZSCH STA 449-F3 Jupiter device and TEM images were taken with a JEOL JSM-207 7600F device. X-ray fluorescence (XRF) measurements were performed on a Rigaku NEXCG device under helium atmosphere and RX5 target.

*3    SBA-15 and KIT-6 mesoporous silica templates*

SBA-15 was synthesized by dissolving 4 g P123 in 120 mL 2 mol/L HCl solution and 30 mL distilled water at RT. Next, the solution was heated to 45 °C and 9.1 mL TEOS was added while vigorously stirring for 5 h. Subsequently, the mixture was aged at 90 °C for 18 h under static conditions. Finally, the precipitate was collected via filtration, dried overnight at RT, and calcined at 550 °C for 6 h in air with a heating rate of 2 °C/min.[2]

KIT-6 was synthesized by dissolving 4 g P123 in 150 mL 0.5 mol/L HCl solution and 4.9 mL butanol at RT. Next the solution was heated to 35 °C and 9.2 mL TEOS was added while vigorous stirring. The mixture was stirred at that temperature for 24 h and subsequently aged at 90 °C for 24 h under static conditions. Finally, the precipitate was collected via filtration, dried overnight at RT and calcined at 550 °C for 6 h in air with a heating rate of 2 °C/min.[3] Low angle X-ray powder diffraction data (XRPD) of SBA-15 and KIT-6 are shown in Figure ESI 1a, indicating that the synthesized KIT-6 mesoporous silica possesses a cubic, interconnected pore system with the $Ia\overline{3}d$-space group symmetry, while SBA-15 shows a $P6mm$ hexagonal ordered pore system. Figure ESI 1b and c show respectively the N$_2$ sorption isotherms and the BJH pore size distribution plot of both silica templates. The pore diameter according to the BJH model was calculated on the desorption branch of the isotherm. Typically, SBA-15 had pores with a diameter of approximately 7.5 nm. The BET surface area and pore volume were typically between 600 and 1000 m$^2$/g and 0.7 to 1 mL/g, respectively. The pore diameter of KIT-6 was generally approximately 7 nm, while the BET surface area ranged from 700 to 900 m$^2$/g and the pore volume from 0.8 to 1.2 mL/g. SBA-15 was used as template material for the synthesis of Bi nanowires for TEM analysis, while KIT-6 mesoporous silica was used as template for the preparation of mesoporous Bi for N$_2$ sorption analysis and for the electric measurements of a Bi – SiO$_2$ (KIT-6) nanowire composite.

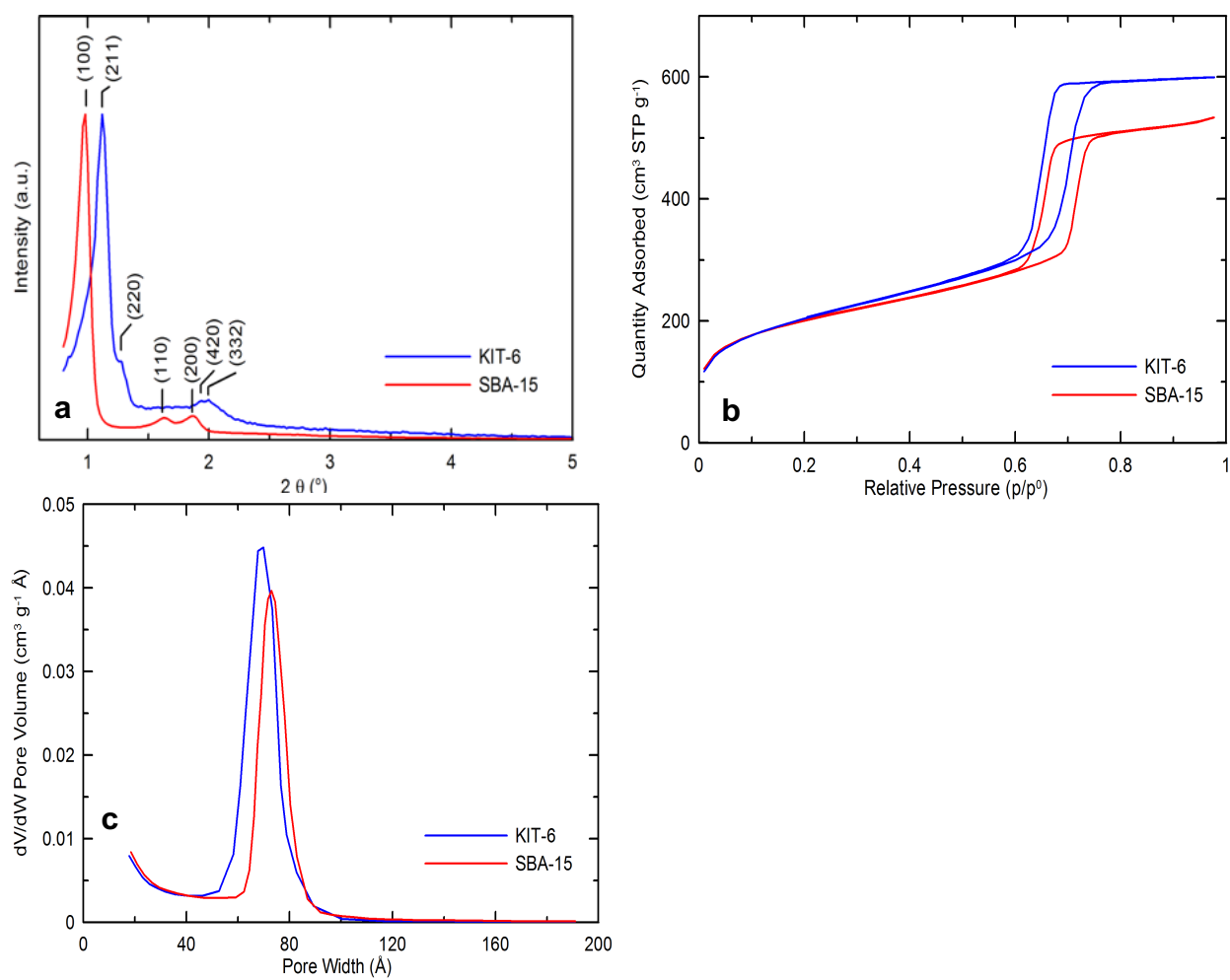

*Figure ESI 1. (a) Low angle XRD diffractogram, (b) N$_2$ sorption isotherms and (c) BJH pore size distribution plot of SBA-15 and KIT-6 mesoporous silica.*

## 4 Continuous-feed nanocasting process

Upon replicating the pore structures of SBA-15 and KIT-6 mesoporous silica, nanowire-like structures and 3D interconnected "nanowire" networks, respectively, were obtained. A schematic representation of the nanocasting process is shown in Figure ESI 2.

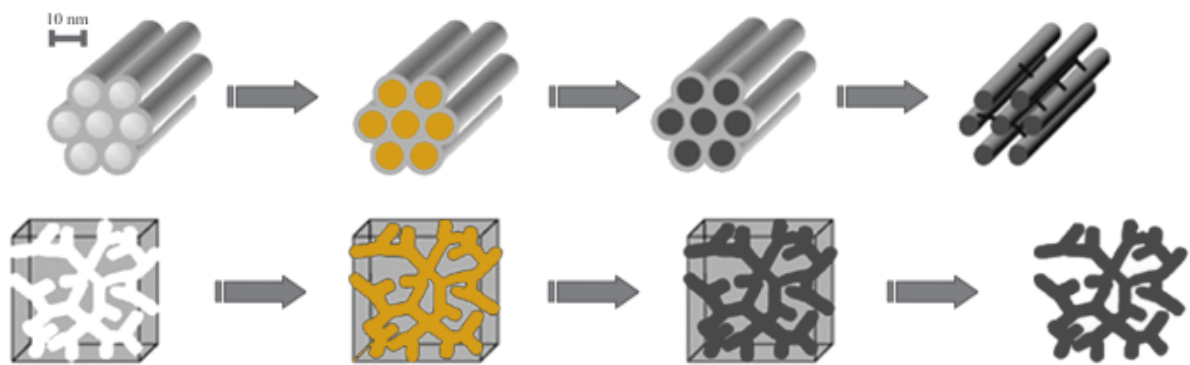

*Figure ESI 2. Schematic representation of nanocasting process for (top) SBA-15 and (bottom) KIT-6 mesoporous silica. First, a bismuth precursor solution is impregnated, resulting in the deposition of $BiCl_3$ salt inside the pores. Second, the $BiCl_3$-$SiO_2$ nanocomposite powder is reduced to Bi – $SiO_2$. Finally, the nanocomposite powder is either sintered, forming a bulk nanowire composite sample, or the silica is chemically etched, allowing TEM imaging or analysis of the mesoporosity.*

The impregnation procedure developed in this research is based on the double solvent method[4] and incipient wetness technique[5]. The example in Figure ESI 3 explains both methods. In Figure ESI 3a-c, a 0.5 mL titania peroxo solution was added to 50 mL toluene. Due to the immiscibility of both solvents, a two-phase system forms. Upon the addition of 1 g SBA-15 with a pore volume of 1 $cm^3$/g, the coloured solution infiltrates the silica template. Since the volume of precursor solution is lower than the total pore volume of the mesoporous silica, all precursor was impregnated inside the template through capillary impregnation. The term "incipient wetness" indicates that the amount of precursor solution added to the template was less or equal to its pore volume. Consequently, deposition of precursor outside the template is prevented. It can be understood from the example that the template's pores are filled with a diluted precursor solution. Consequently, the amount of metal salt deposited inside the pores depends on the metal concentration of the precursor solution.

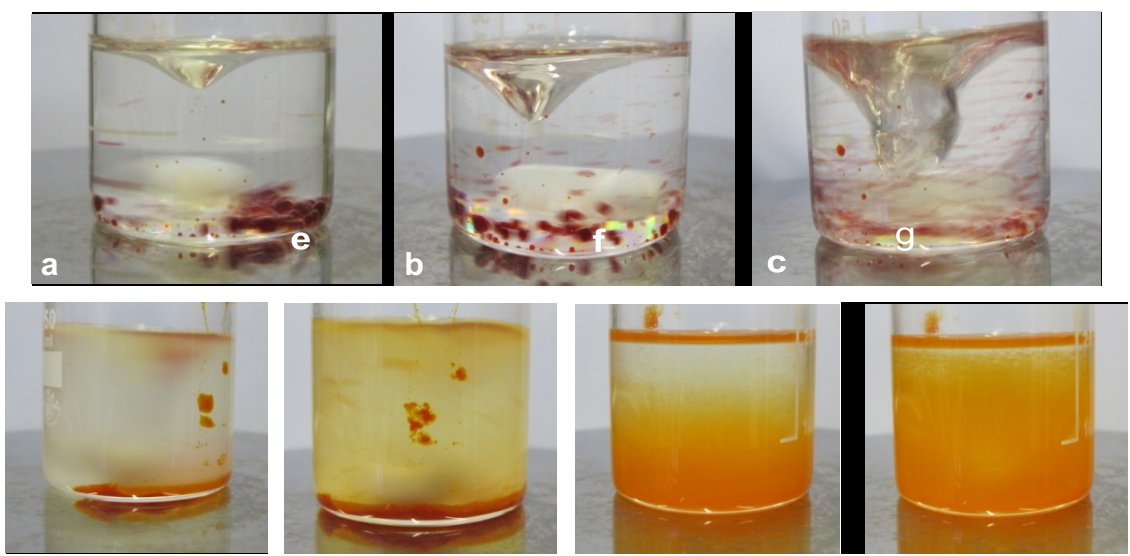

*Figure ESI 3. Visual representation of "double solvents" impregnation method: (a-c) a titanium peroxo complex in diluted hydrochloric acid (dark orange aqueous precursor solution) in toluene while vigorously stirring. (d-g) Titania precursor solution in a dispersion of mesoporous silica in toluene after 0, 2, 5, and 10 min stirring, respectively.*

In contrast to the double solvent method, the process presented here makes use of refluxing non-polar solvent with a boiling point higher than water, such as toluene, xylene, or n-octane. The polar precursor solution is added continuously and the total volume added is many times larger than the template's pore volume. By means of a Dean-Stark separator, all water from the metal precursor solution is removed from the system while the metal solution immobilizes inside the template's pores, which enables the complete filling of the template with the metal salt by impregnation of a diluted precursor solution. A PFA round bottom flask was used since its hydrophobic properties prevent the wetting of the internal surface or the reaction vessel. The low addition rate of the aqueous precursor solution to the dispersed silica was chosen to ensure that the amount of precursor solution added was always lower than the total pore volume of the template.

Both mesoporous silica templates, SBA-15 and KIT-6, were impregnated using the same procedure described below. Prior to the impregnation of either of the silica templates, the pore volume was measured by means of $N_2$ sorption analysis to derive the mount of metal salt the pores can contain. The polar precursor solution is defined as an aqueous solution of the precursor salts $BiCl_3$ or $Bi(NO_3)_3 \cdot 5H_2O$ dissolved in a mixture of precursor solvents, HCl, $H_2O$, methanol (MeOH) and/or formic acid. The non-polar solvents toluene, xylene, or n-octane are used in such combination that they are immiscible with the polar precursor solution. The different impregnations performed in this study are listed in Table ESI 1.

*Table ESI 1. List of impregnation conditions.*

| Precursor salt | Precursor solvent | Non-polar solvent | Precursor addition speed (mL/h) | Temperature heater (°C) |
|---|---|---|---|---|
| $Bi(NO_3)_3 \cdot 5H_2O$ | Water, $HNO_3$ | toluene, xylene, n-octane | 4 | 140-145 |
| $BiCl_3$ | Water, HCl, formic acid (FA), methanol | n-octane | 4 | 160-165 |

The impregnation setup depicted in Figure ESI 4 was used for the synthesis of bismuth nanowires and nanocomposite structures. An aluminium heat exchanger was used for efficient heating of the PFA round bottom flask. A syringe pump was used to inject the precursor solution in a controlled manner, whereas the temperature of the heater controlled the evaporation rate of the aqueous solution. Note that the simplicity of the system did not allow us to monitor the rate of water removal by means of the Dean-Stark separator.

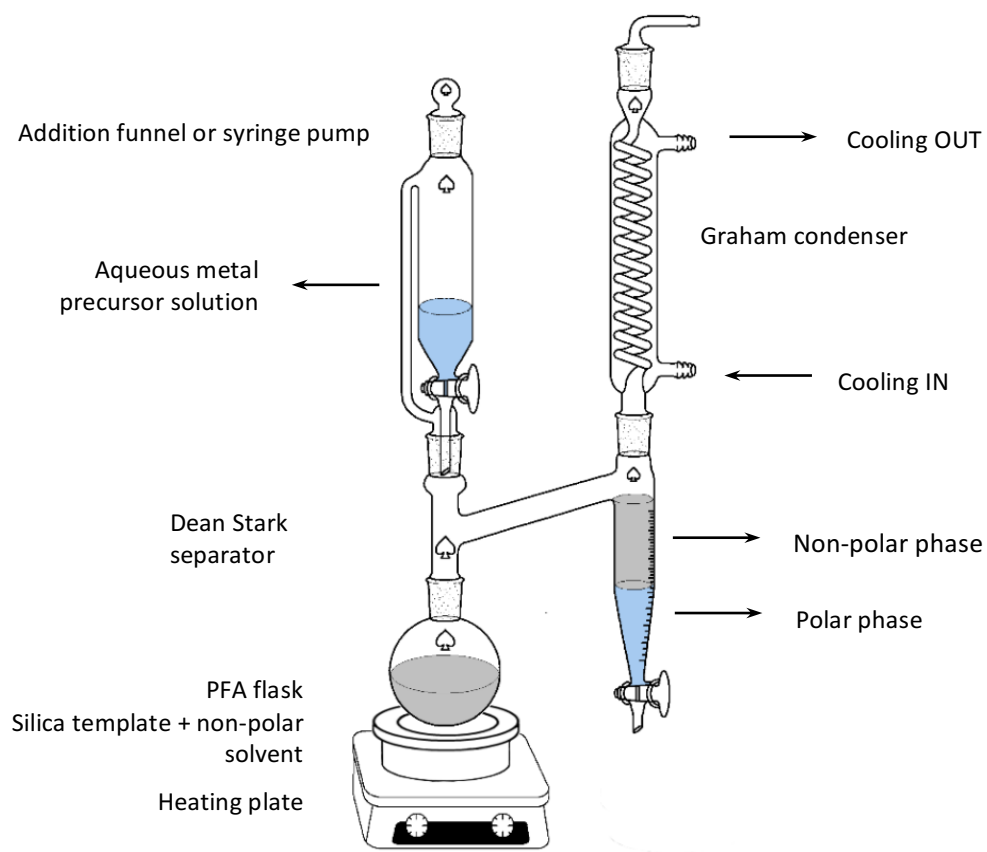

*Figure ESI 4. Schematic representation of the continuous-feed nanocasting setup for the impregnation of a polar precursor solution into a mesoporous silica template dispersed in a high-boiling, non-polar solvent. The polar precursor solution is added to the PFA reaction flask at a constant rate and migrates into the pores. Meanwhile, the polar precursor solvent is removed selectively from the system by means of the Dean-Stark separator. The precursor salt or its decomposition product is deposited inside the template's pores. The refluxing solvents boil off and condense in the Graham condenser. Due to the higher density and the immiscibility of both solvents, the aqueous phase is selectively collected in the right arm of the Dean Stark, while the non-polar solvent returns to the system.*

Since the bismuth precursor salt $BiCl_3$ hydrolyses in water, an acidified aqueous precursor was prepared. However, we observed from Figure ESI 5 that the metal salt decomposed to BiOCl during the impregnation of a 30 w% HCl and 60 mL methanol, both with and without the addition of 15 mL formic acid in n-octane (bp. 125 °C) as non-polar solvent.

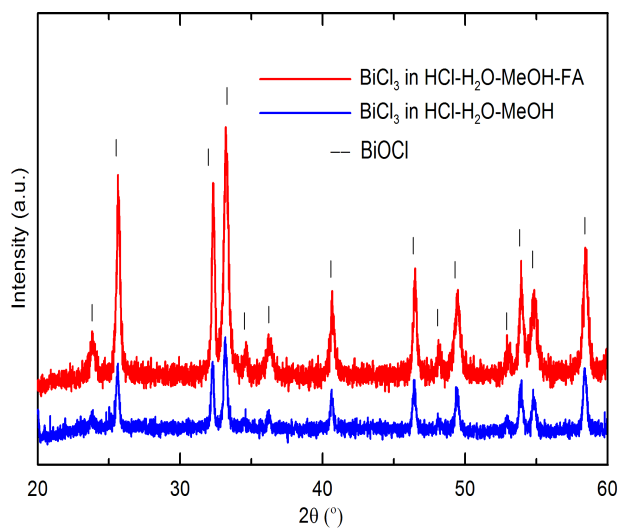

*Figure ESI 5. X-ray powder diffraction pattern of SBA-15 mesoporous silica powder impregnated with $BiCl_3$ dissolved in 15 mL 30 w% HCl and 60 mL methanol, with or without 15 mL formic acid (FA), while octane was used as non-polar solvent. It can be seen that the metal salt decomposed to BiOCl as the solvent was removed during the impregnation process.*

Since the molecular weight of the precursor salt is typically higher and the density lower than the desired product, volume constriction of the impregnated precursor during the conversion of, e.g., a metal salt to its oxide or metallic form is inevitable. However, the loading efficiency of a certain material can be enhanced if the difference in molecular weight between the decomposed precursor and the target product is reduced. The decomposition of $BiCl_3$ to BiOCl during the impregnation is an example how the pore loading can be enhanced. Table ESI 2 shows the theoretical maximum pore loading efficiencies of $Bi(NO_3)_3 \cdot 5H_2O$ and $BiCl_3$ depending on their decomposed precursor when they are converted to their metallic form.

*Table ESI 2. Theoretical maximum pore loading efficiencies with metallic bismuth depending on the precursor salt used and its decomposition product during the impregnation.*

| Precursor salt | Decomposed precursor | Target product | Pore loading efficiency (%) |
| --- | --- | --- | --- |
| $Bi(NO_3)_3 \cdot 5H_2O$ | None ($Bi(NO_3)_3 \cdot 5H_2O$) | Bi | 12.5 |
| $Bi(NO_3)_3 \cdot 5H_2O$ | $BiONO_3$ | Bi | 36.7 |
| $BiCl_3$ | None ($BiCl_3$) | Bi | 32.2 |
| $BiCl_3$ | BiOCl | Bi | 60.5 |

To obtain a better understanding of the decomposition process of the precursor salt during the impregnation, TGA measurements were performed. The thermal decomposition of $Bi(NO_3)_3 \cdot 5H_2O$ and $BiCl_3$ in air is shown in Figure ESI 6. It can be calculated from the analysed mass loss that the nitrate salt lost 4.4 hydrate water molecules when heated to 108 °C, the boiling temperature of toluene, while a mass loss equivalent to 6 water molecules was obtained at 125 °C, the boiling temperature of n-octane. The latter suggests that on top of the removal of all water molecules of $Bi(NO_3)_3 \cdot 5H_2O$, the nitrates also started to decompose. Dry BiCl$_3$, on the other hand, started to decompose only above 250 °C, while a melt was formed at approximately 228 °C (endothermic peak in DTA signal). We believe the drastic weight loss of BiCl$_3$ between approximately 200 and 330 °C was due to sublimation of the salt, which could be suppressed by using humidified air. The decomposition of the precursor salt during impregnation is possible due to the high temperature used during the impregnation, namely, refluxing toluene, n-octane, or xylene. Due to removal of the acidic precursor solvent during the impregnation, the BiCl$_3$ metal salt hydrolysed to BiOCl, while Bi(NO$_3$)$_3$.5H$_2$O decomposed through the loss of hydrate waters. The extent of these effects depends on the boiling temperature of the non-polar solvent.

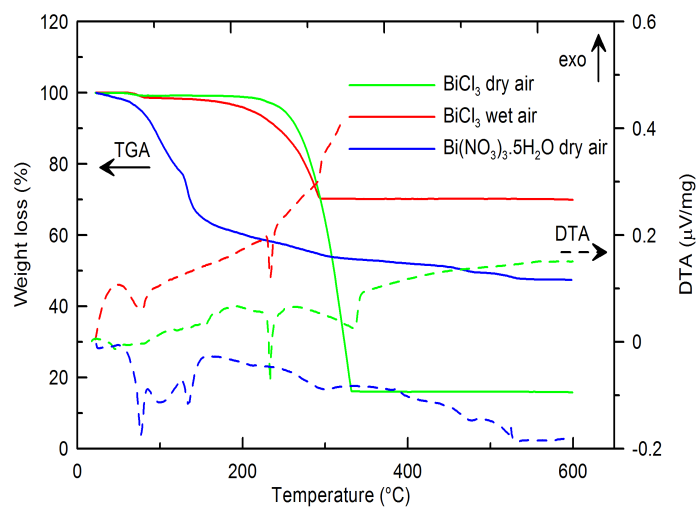

*Figure ESI 6. Thermogravimetric analysis of Bi(NO$_3$)$_3$.5H$_2$O and BiCl$_3$ in humidified and dry air at 2 °C/min heating rate. The weight loss due to the loss of hydrate waters is clear from the TGA signal of Bi(NO$_3$)$_3$.5H$_2$O. The TGA signal of BiCl$_3$ shows a drastic weight loss in dry air, whereas the weight loss was suppressed in humidified air. The endothermic peak on the DTA signal of BiCl$_3$ shows the formation of a melt at 228 °C.*

## 5   Reduction of BiCl$_3$ – SiO$_2$ nano composite powder

A sealable tube flow furnace (Figure ESI 7) was used for the reduction of BiCl$_3$/Bi(NO$_3$)$_3$ – silica composite powder. Approximately 0.5 g of sample was placed in a ceramic crucible and placed in the centre of the furnace. Different reducing agents were tested to reduce Bi(NO$_3$)$_3$.5H$_2$O and BiCl$_3$ to metallic bismuth. However, the low melting point of metallic Bi, 271 °C, was a major limitation on the applicability of conventional reducing gas, such as Ar-5% H$_2$. The reduction was performed by bubbling an Ar-5 % H$_2$ gas flow through a solution of 98 % hydrazine monohydrate, transferring hydrazine vapour in the furnace. The reduction was performed between 220 and 265 °C for 10 h.

Vandaele et al.

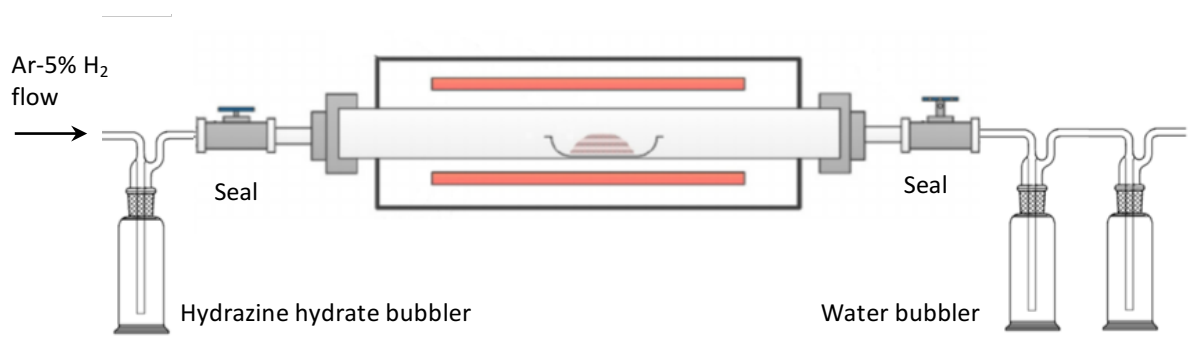

*Figure ESI 7. Sealable horizontal flow furnace for the reduction of Bi(NO$_3$)$_3$/BiCl$_3$ – SiO$_2$ nanocomposite powder. The glass furnace tube can slide out of the furnace and transferred to the glovebox for further manipulation of the powder.*

Bulk Bi$_2$O$_3$ and BiCl$_3$ powder were used to investigate the influence of different atmospheres on their reduction to metallic Bi. ZnO was used as internal standard to perform Rietveld refinement and calculate the fraction of crystalline vs amorphous Bi, as well as the amount of secondary phases. The heat treatments were conducted in a tubular furnace in a flow of Ar – 5% H$_2$ gas, either pure or bubbled through a solution of hydrazine monohydrate (N$_2$H$_4$·H$_2$O). A flow rate of 20 mL/min was used and the samples were reduced for 10 h at 220 °C, 250 °C, and 265 °C. Hydrazine was used here as it is a powerful and clean reducing agent, since all decomposition products are gaseous.[6,7] However, a minimum concentration of 64 % hydrazine solution is required to obtain sufficient reducing power. Namely, the reducing capacity reduces quickly as the solution becomes more diluted. X-ray diffraction (XRD) patterns of BiCl$_3$ reduced at different temperatures in a flow of Ar - 5% H$_2$ and in the presence of N$_2$H$_4$ vapour are depicted in Figure ESI 8.

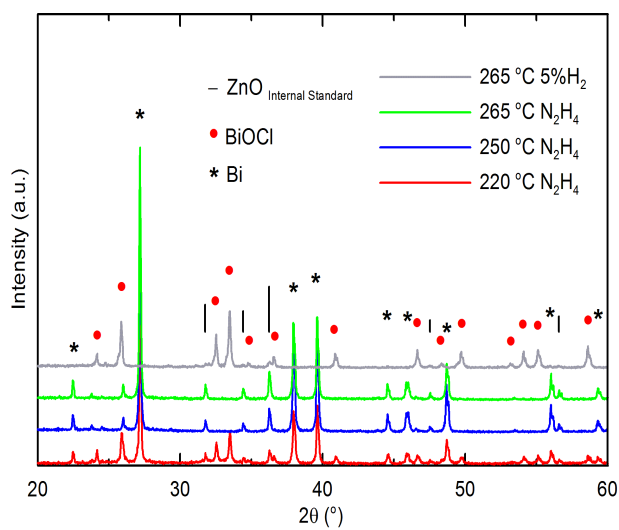

*Figure ESI 8. X-ray diffraction patterns of BiCl$_3$ reduced at different temperatures in Ar - 5% H$_2$ and in the presence of N$_2$H$_4$ vapour. The diffraction peaks of the internal standard ZnO, which was used to quantify the amount of crystalline Bi and secondary phase BiOCl.*

The Rietveld refining data calculated from the recorded X-ray diffraction pattern is reported in Table ESI 3. It is shows that a minimum temperature of 250 °C and a reduction time of 10 h is required to reduce $BiCl_3$ in $N_2H_4$ vapour to metallic Bi. In the absence of hydrazine vapour, very little Bi phase was formed when reduced at 265 °C in Ar-5% $H_2$.

*Table ESI 3. Rietveld refinement data of $BiCl_3$ powder reduced at different temperatures and atmospheres for 10 h. The data shows that the presence of $N_2H_4$ vapour was crucial to reduce $BiCl_3$ to metallic Bi. A temperature of 250 °C and a reduction time of 10 h in $N_2H_4$ loaded Ar - 5% $H_2$ was required to reduce $BiCl_3$.*

| Material | Red. Temp. | Atmosphere | Bi | $Bi_2O_3$ | BiOCl | Amorphous |
|---|---|---|---|---|---|---|
| $BiCl_3$ | 220 °C | 5 % $H_2$ + $N_2H_4$ | 32 % | 0 % | 17 % | 51 % |
| $BiCl_3$ | 250 °C | 5 % $H_2$ + $N_2H_4$ | 34 % | 0 % | 0 % | 66 % |
| $BiCl_3$ | 265 °C | 5 % $H_2$ + $N_2H_4$ | 26 % | 0 % | 0 % | 74 % |
| $BiCl_3$ | 265 °C | 5 % $H_2$ | 1 % | 0 % | 44 % | 55 % |

Although a minimum temperature of 250 °C was required to reduce bulk $BiCl_3$ powder with $N_2H_4$, the nanocomposite powders were typically reduced at 220 °C. We believe that the highly exposed surface area facilitated the reduction of $BiCl_3$ – silica nanocomposite powder. Also, we can see from TEM analysis that bismuth leached out of the pore channels upon increasing reduction temperature to 265 °C, as shown in Figure ESI 9. The occurrence of bismuth outside the template can be either by leaching of $BiCl_3$ precursor due to the formation of a melt at 228 °C, or due to the melting of Bi.

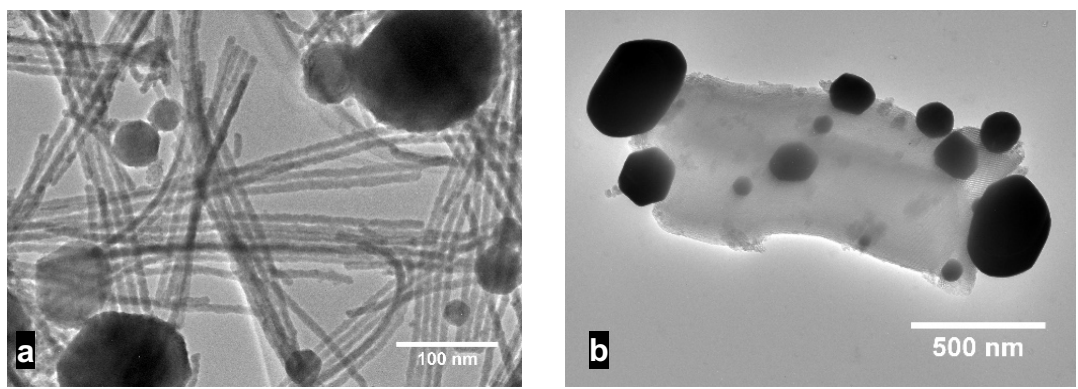

*Figure ESI 9. (a) TEM image of 30 v% Bi loaded in SBA-15 and subsequent leaching of the template. $BiCl_3$ was impregnated dissolved in a 20 v% 7.5 mol/L HCl - MeOH solution, with n-octane as non-polar solvent and subsequently reduced at 220 °C in a hydrazine-loaded Ar-5% $H_2$ gas flow for 10 h. The template was chemically etched with a 1 mol/L NaOH – 0.5 mol/L hydrazine solution for 3 h. (b) TEM image of 30 v% Bi loaded in SBA-15. $BiCl_3$ was impregnated dissolved in a 10 % 7.5 mol/L HCl - MeOH solution, with n-octane as non-polar solvent and subsequently reduced at 265 °C in a hydrazine-loaded Ar - 5% $H_2$ gas flow for 10 h. The light grey area is the silica template, while the black spheres are bismuth spheres which were formed due to leaching of Bi out of the pores and sintering on the template's exterior surface. Note that the Bi nanowires were compromised during TEM sample preparation, depicting oxidized Bi nanowires.*

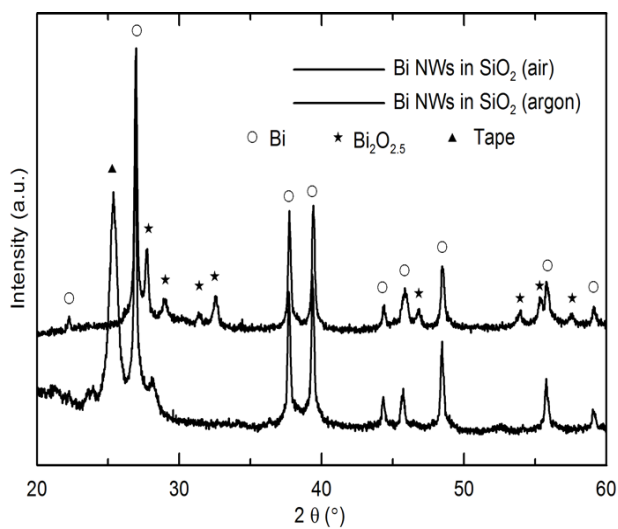

*Figure ESI 10. X-ray powder diffraction (XRPD) data of Bi nanowires embedded in KIT-6 mesoporous silica template. (top) Measurement performed under argon atmosphere by applying tape on sample prepared in glovebox. (bottom) XRPD pattern of same sample without tape shows presence of $Bi_2O_{2.5}$ reflections. The sharp reflections indicate single-crystal nanowires. However, this has not yet been confirmed by HR TEM, as oxidation occurs during TEM sample preparation*

After completing the nanocasting process, the replica was separated from the silica template by etching the silica using a 1 mol/L sodium hydroxide solution for 3 h. The nanostructures were recovered through centrifugal separation and washed 4 times with water and ethanol. The $N_2$ sorption isotherms of mesoporous Bi replicated from a KIT-6 mesoporous silica is shown in Figure ESI 11. The samples were prepared by impregnating a $BiCl_3$ precursor solution in refluxing n-octane or the impregnation of $Bi(NO_3)_3$ in toluene. The composite powders were reduced at 230 °C for 12 h in hydrazine vapour and subsequently the silica template was removed by chemical etching. The powders were dried for 24 h at 140 °C prior to analysis. A BET surface area of 20 and 13 $cm^3$/g were obtained for the samples prepared with respectively $Bi(NO_3)_3$ and $BiCl_3$. No long-range ordering of the pores was observed in low-angle XRD.

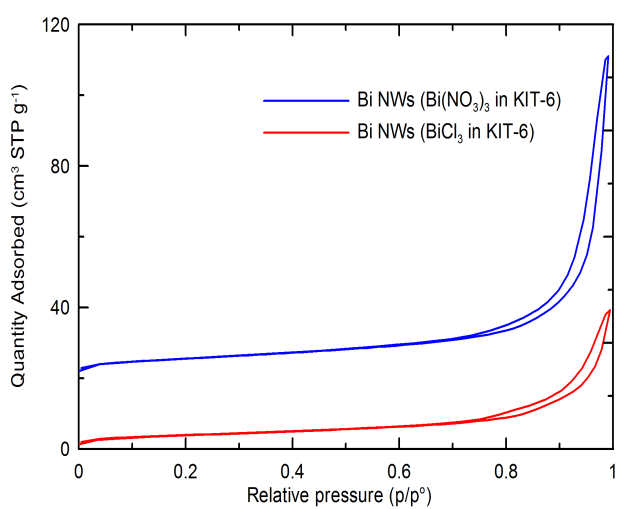

*Figure ESI 11. $N_2$ sorption isotherms of mesoporous Bi replicated from KIT-6 mesoporous silica, by the impregnation of $Bi(NO_3)_3$ or $BiCl_3$ precursor salt, reduction at 230 °C and template removal in 1 mol/L NaOH solution.*

## 6  Synthesis of Bi – SiO$_2$ nano composite powder

KIT-6 mesoporous silica was used as template for the synthesis of Bi nanowire composites. The replicated structure after chemically etching the silica template is shown in Figure ESI 12.  The nanowire composite powder was sintered at 230 °C for 20 min in a vacuum under a uniaxial pressure of 50 MPa and subsequently cut for further analysis. Note that all steps of the sample preparation and the electrical measurements were performed under inert atmosphere.

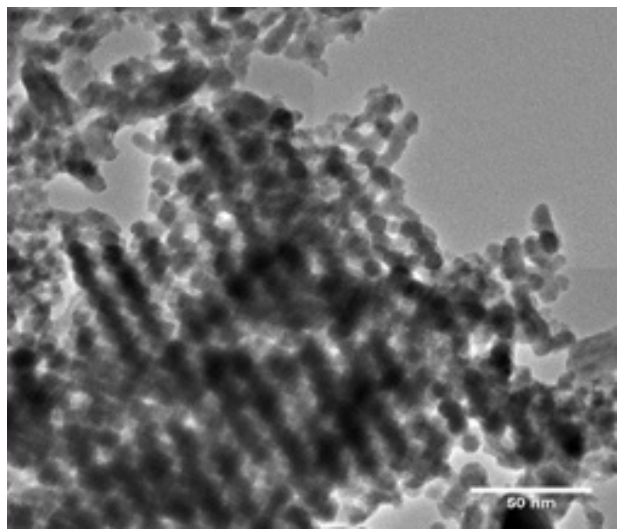

*Figure ESI 12. TEM images of Bi nanowire arrays replicated from KIT-6 mesoporous. The silica template was impregnated with a BiCl$_3$ precursor in 10 w% HCl – 80 v% MeOH solution, reduced at 220 °C in N$_2$H$_4$ vapour for 12 h and subsequently subjected to 1 mol/L NaOH etching solution to remove the silica template. Note that the interconnected Bi nanowires were compromised during TEM sample preparation, depicting oxidized Bi nanowires.*